\documentclass[twocolumn,pre,showpacs,amsmath]{revtex4-1}
\usepackage{epsfig,amsfonts,amsbsy}
\usepackage[usenames, svgnames]{xcolor}
\usepackage{bm}

\newcommand{\ket}[1]{\left|#1 \right\rangle}

\newcommand{\state}[1]{\left \vert \left. #1 \right\rangle   \right.}

\newcommand{\non}{\nonumber}

\newcommand{\rmL}{{\rm L}}
\newcommand{\rmR}{{\rm R}}

\newcommand{\sectionprl}[1]{{\par\it #1.---}}
\usepackage{ulem}
\usepackage[mathlines]{lineno}
\usepackage{colortbl}
\usepackage{tabularx}
\usepackage{verbatim}
\usepackage{multirow}
\usepackage{umoline}
\usepackage[hyperindex,breaklinks]{hyperref}
\hypersetup{
     colorlinks=true,       		
     linkcolor=Navy,          	
     citecolor=Navy,            
     filecolor=Navy,      		
     urlcolor=Navy,           	
    runcolor=cyan,
 }

\begin{document}
\title{Many-body scar state intrinsic to periodically driven system: Rigorous results}
\author{Sho Sugiura$^{1}$ ,Tomotaka Kuwahara$^{2,3,4}$ and Keiji Saito$^{5}$}
\affiliation{$^{1}$Department of Physics, Harvard University, Cambridge, MA 02138, USA}
\affiliation{$^{2}$Mathematical Science Team, RIKEN Center for Advanced Intelligence Project (AIP),1-4-1 Nihonbashi, Chuo-ku, Tokyo 103-0027, Japan}
\affiliation{$^{3}$Department of Mathematics, Faculty of Science and Technology, Keio University, 3-14-1 Hiyoshi, Kouhoku-ku, Yokohama 223-8522, Japan}
\affiliation{$^{4}$Interdisciplinary Theoretical \& Mathematical Sciences Program (iTHEMS) RIKEN 2-1, Hirosawa, Wako, Saitama 351-0198, Japan}
\affiliation{$^{4}$Department of Physics, Keio University, Yokohama 223-8522, Japan}

\date{\today}

\begin{abstract}
  The violation of the Floquet version of eigenstate thermalization hypothesis is systematically discussed with realistic Hamiltonians. Our model is based on the PXP type interactions without disorder. We exactly prove the existence of many-body scar states in the Floquet eigenstates, by showing the explicit expressions of the wave functions. Using the underlying physical mechanism, various driven Hamiltonians with Floquet-scar states can be systematically engineered.   
\end{abstract}
\maketitle

In the past few decades, significant progress have been made in the in-depth understanding of the thermalization phenomenon in isolated systems \cite{popescu2006entanglement,Goldstein2006,Goold_2016,Yukalov_2011,doi:10.1146/annurev-conmatphys-031214-014726,doi:10.1080/00018732.2016.1198134,Gogolin_2016,Mori_2018}. Thermalization is a fundamental phenomenon in physics, which is directly connected to {\it Arrow of time} in the sense of thermodynamics. In addition, it provides the underlying fundamental mechanism to validate the framework of statistical mechanics.

The eigenstate thermalization hypothesis (ETH) is one of the most important keywords in this subject for static systems, because if it holds the thermodynamic property in the isolated systems is consistently explained \cite{Reimann_2015,Deutsch_2018,ETHDeutch1991,ETHSrednicki1994}. The ETH states that any single eigenstate is {\it thermalized} in the sense that an expectation value of any local observable is equal to the value calculated by the canonical ensemble with the corresponding temperature. When a periodic driving is applied to the system, the system generally heats up. In this case, the standard ETH is replaced by another hypothesis known as the Floquet ETH, which states that any single Floquet eigenstate is thermalized with an infinite temperature~\cite{dalessio13,dalessio14,lazarides14,kuwahara16,mori16}. Both the ETH and the Floquet ETH have been intensively studied, and the affirmative results of many numerical simulations corroborate their validity, as long as the system is nonintegrable \cite{kim2014testing,mondaini2016eigenstate,mondaini2017eigenstate,dalessio14,mori16,Asmi2019}. However, exceptions exist for both the ETH and the Floquet ETH. The most common example is the many-body localization, which is a phenomenon driven by disorder \cite{doi:10.1146/annurev-conmatphys-031214-014726,ponte15_1,ponte15_2}. In the system exhibiting such phenomena, essentially all eigenstates are nonthermal states, which are protected by an extensive number of emergent local integrals of motion.

Recently, a new type of violation of the ETH has been found in the PXP model in the framework of {\it many-body scar state} \cite{turner18, TurnerScar2018_2,AKLT, WenScar2019,VedikaScar2019,JamesConfinement2019,SoonwonScar2019}. The PXP model is an effective model derived from the transverse Ising model that describes the experimental setup of a chain of Rydberg atoms \cite{bernien}. The many-body scar states have been numerically proposed as nonthermal eigenstates which consistently explain the long-time oscillations observed in experiments \cite{bernien, WenScar2019, SoonwonScar2019}. In addition, Lin and Motrunich found the explicit expressions for several nonthermal eigenstates using the matrix product form \cite{LM}. This work has rigorously ensured the existence of many-body scars in this system and has made their structure clearer. An intriguing nature is the absence of a local integral of motion, which is in stark contrast to the conventional theories on the breakdown of the thermalization such as the many-body localized and the integrable systems.

Very recently, a similar violation of the Floquet ETH has been numerically reported in a random unitary time-evolution \cite{pai19,VedikaNandkishoreFracton19} that models the Fracton dynamics~\cite{Nandkishore:2019aa, PaiPRX19}. In this study, the unitary dynamics was randomly chosen, while the translational symmetry and several conservation laws were retained. The study numerically indicated that the Floquet operator has {\it Floquet many-body scar states} that imply nonthermal eigenstates of the Floquet operator. Because this is an important numerical indication, several questions should follow: (i) Can such a scar exist in a systematic Hamiltonian? If yes, what is the possible mechanism? (ii) Can we construct exact expressions for Floquet scar states, as was done for the scar states in the PXP model? Addressing these questions is indispensable for an in-depth understanding of the general structure of scar states, as well as for future experimental realization. 

In this paper, we rigorously discuss the violation of the Floquet ETH in a systematic Hamiltonian. Our model is based on the PXP type interactions without disorder. Then, we rigorously prove the existence of the many-body Floquet scar states by deriving the explicit expressions of the eigenstates. Through the derivation of the wave functions, underlying mechanisms to have the Floquet scar states is clarified in this model. In addition, this mechanism enables us to engineer various Hamiltonians with Floquet scar states. Since our model is based on PXP type of interactions \cite{bernien,YaoFSPT17}, simple cases of our Floquet scar states are experimentally feasible.
By contrast, other eigenstates should satisfy the Floquet ETH. 
We numerically demonstrate that a quantum state usually relaxes to a state with infinite temperature in our system, while our Floquet scar state exhibits persistent oscillations which never decay. 

\sectionprl{Floquet-intrinsic many-body scar state}
\begin{figure}
		\includegraphics[width= 0.45 \textwidth]{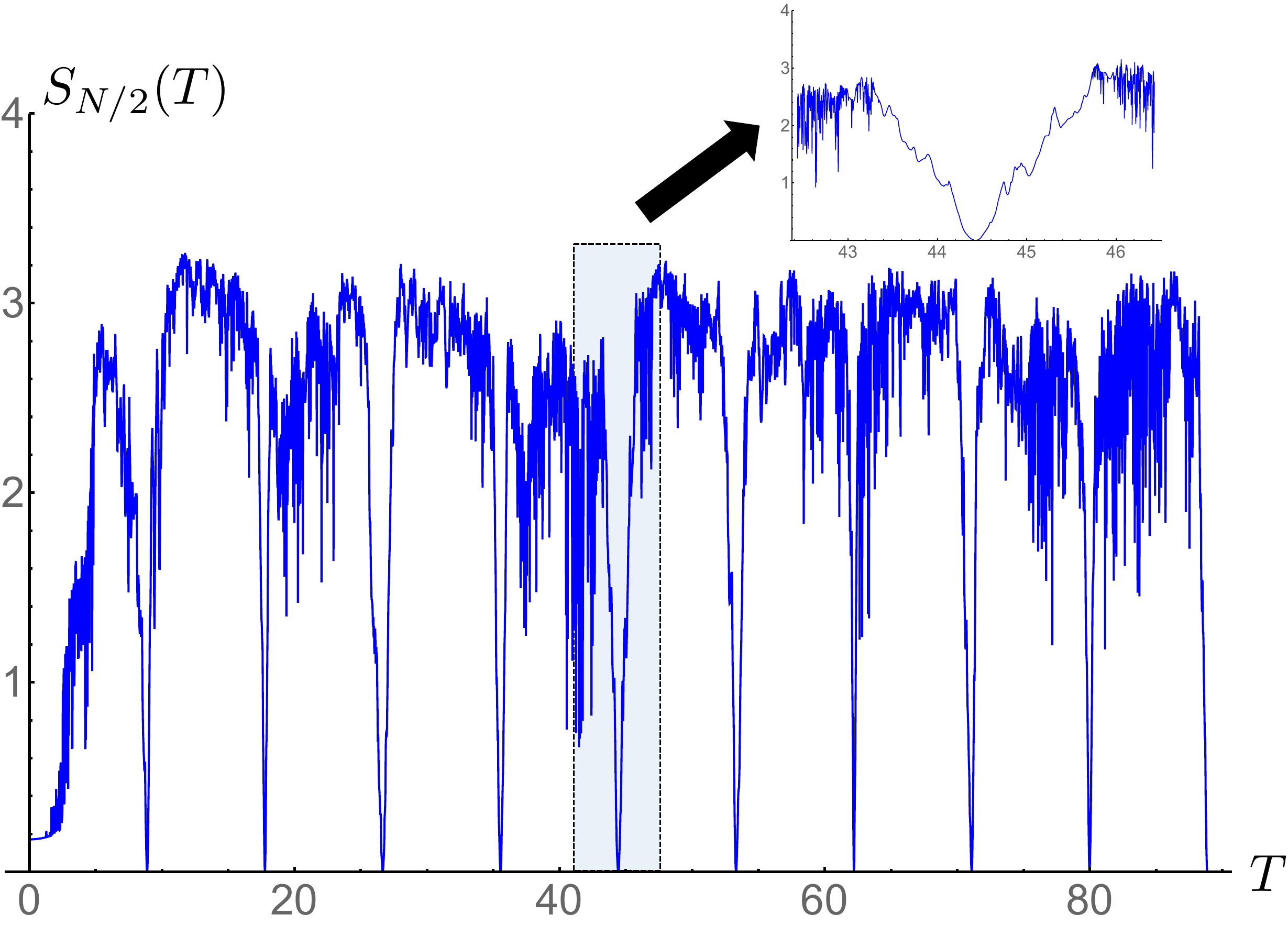}
		\caption{$T$-dependence of the minimum values of the entanglement entropies for the Floquet operator $F$ 
		with $N=16$. For $T=0.004m \cdot 2\pi/\sqrt{2}$ with $m=1,2,\ldots, 5000$, we calculate the entanglement entropy $S$ for all the eigenstates of the Floquet operator $F$ and select the minimum value among them.
                }
		\label{tdepe}
\end{figure}
Let $H(t)$ be a time-dependent many-body Hamiltonian, which is periodic in time with the period $T$. The Floquet operator for a single period is given by
\begin{align}
  F &= {\cal T} \exp \left( -i \int_0^T dt H(t) \right) \, ,
\end{align}
where ${\cal T}$ is a time-ordering operator. We set $\hbar$ to be unity. For simplicity, we consider the following time-dependence in the Hamiltonian: 
\begin{eqnarray}
  H(t) &=& \left\{ 
           \begin{array}{ll}
             H_1 & 0 \le t < T/2 \,   \\
                   H_2 & T/2 \le t < T \,  
             \end{array} \, .  \right.
\end{eqnarray}
Hereinafter, we assume that $H_1$ and $H_2$ do {\it not} commute with each other and that both of the Hamiltonians are nonintegrable. The Floquet operator is now simply written as $F = e^{-i H_2 T/2} e^{-i H_1 T/2} $.  According to the Floquet ETH, the Floquet Hamiltonian $H_F$ defined from the relation $e^{-i H_F T} = F$ is generally a random Hamiltonian, whose eigenstates are the states with an infinite temperature \footnote{When $H_F$ has symmetries and corresponding conserved quantities, we divide the Hilbert space into the sectors of the conserved quantities and look into one of the sectors which have exponentially large dimension.}. 
Although explicitly describing the Floquet Hamiltonian is difficult, the Floquet Hamiltonian is generally thought to be far from an integrable Hamiltonian.

To make our objective more explicit, we classify the possible Floquet scar states into two. The first is a trivial case where we have the simultaneous eigenstates of $H_1$ and $H_2$; such states automatically become the eigenstates of the Floquet operator, 
and can be demostrated, e.g., with a frustration-free Hamiltonian. See the footnote 
\footnote{This class of the Floquet scar can be easily demonstrated when we employ the frustration-free Hamiltonian, where the ground state is a simultaneous eigenstate of all local interaction terms. For instance, we can consider the decomposition of the AKLT Hamiltonian. Let $h_j$ be $h_j={\bm S}_{j} \cdot {\bm S}_{j+1} + \left({\bm S}_{j}\cdot {\bm S}_{j+1} \right)^2 /3 $ which is the interaction term between the sites $j$ and $j+1$, where ${\bm S}$ is a spin-1 operator. Then for even size $N$, we set ${H}_1 =\sum_{j=1}^{{N / 2}} {h}_{j}$ and ${H}_2 =\sum_{j=N/2+1}^{N } {h}_{j}$. Note that $H_1$ and $H_2$ are uncommutable to each other, and those are nonintegrable. Frustration-free leads to that the ground state for $H_1+H_2$ is also eigenstate of $H_1$ and $H_2$.}, 
for an explicit example using the AKLT Hamiltonian \cite{AKLT1987}. 
The second is a more nontrivial case, which is investigated in this study. In this class, the Floquet scar states are the eigenstates of ${F}$, but not of $H_j \, (j=1,2)$. To distinguish this class of scars from the first class, we term them as {\it Floquet-intrinsic many-body scar state} (FMS). We discuss the existence of the FMS, and the possible mechanisms in this letter. 

\sectionprl{Model and numerical demonstration}
\begin{figure}
		\includegraphics[width= 0.45 \textwidth]{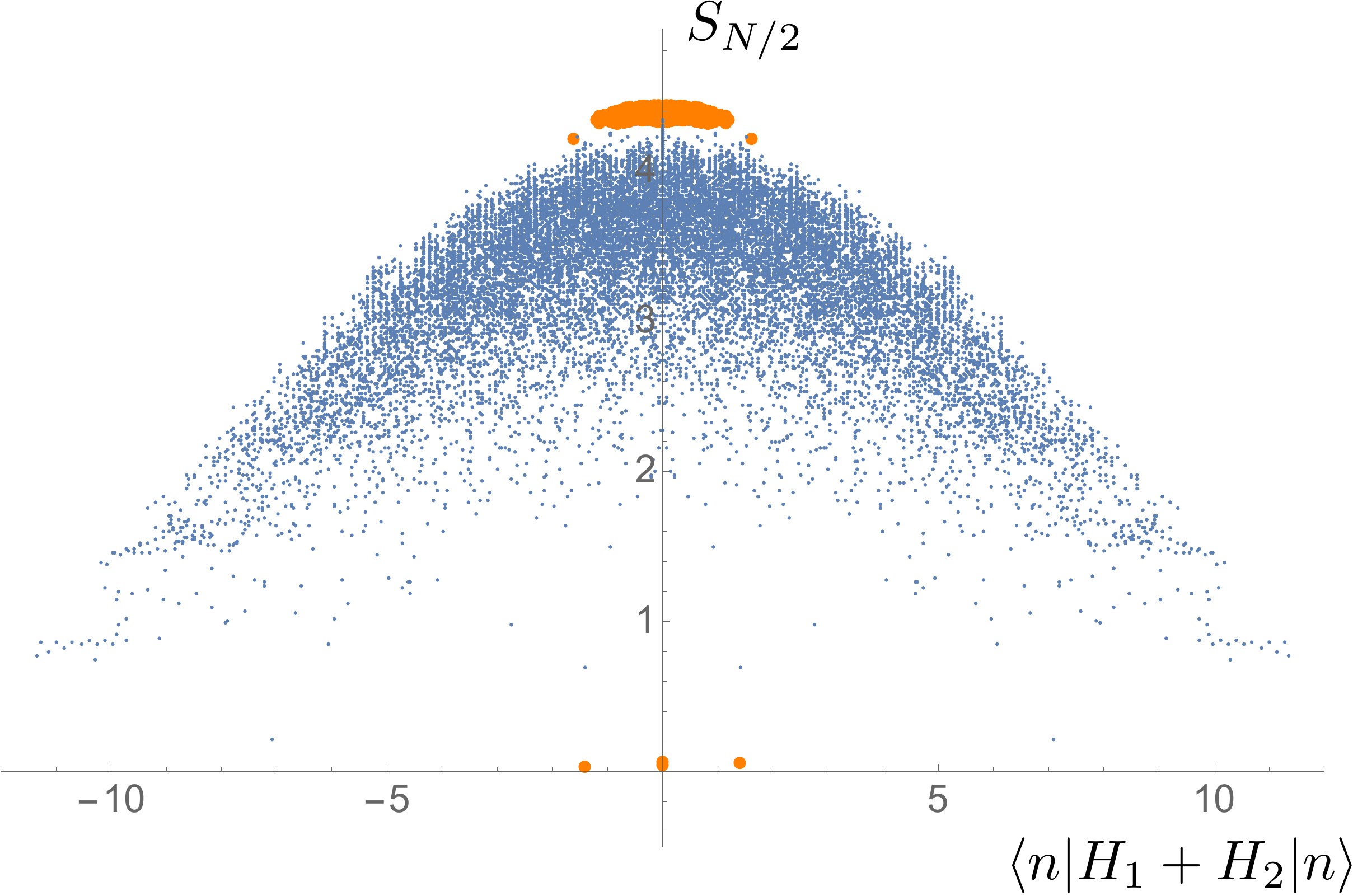}
		\caption{
		Non-trivial Floquet many-body scar states and other energy eigenstates. We calculate all the eigenstates of the Floquet operator for $T=10\sqrt{2}\pi$ with $N=20$ and compare the properties with the static PXP Hamiltonian.
	The horizontal axis and the vertical axis denote the energy expectation with respect to the PXP Hamiltonian $H_1+H_2$ ($\state{n}$ is either an eigenstate of $H_1+H_"$ or Floquet eigenstate of F), and the vertical axis denotes the entanglement entropy, respectively. We observe four eigenstates with zero entanglement with energies $\pm \sqrt{2}$ and $0$, where two scar eigenstates have zero energy.
		}
		\label{fig entanglement}
\end{figure}
We construct a model to investigate the FMS; we use a time-dependent version of the PXP model in this study. 
We set the following combination of Hamiltonians:
\begin{align}
	H_1 &=\sum_{j=2}^{{N / 2}} {P}_{j-1}{X}_{j}{P}_{j+1} +{V}_1 \, , \label{Ham L}\\
	H_2 &=\sum_{j={N / 2}+1}^{N-1} {P}_{j-1}{X}_{j}{P}_{j+1}+{V}_2 \, ,\label{Ham R} 
\end{align}
where $N$ represents the size of the system, which is an even number. The operators ${X}_j$, ${Y}_j$, and ${Z}_j$ are the $x$, $y$, and $z$ components of the Pauli operators, respectively, at the site $j$. Let 
$\ket{\uparrow}_j$ and $\ket{\downarrow}_j$ be the eigenstates of ${Z}_j$ with the eigenvalues $+1$ and $-1$, respectively. The operator ${P}_j$ is a projection operator onto a down spin state at the site $j$, i.e., $ {P}_j := ({1}-{Z}_j)/2$. We consider a constrained Hilbert space without any adjacent up states, e.g., $\ket{\uparrow}_j \ket{\uparrow}_{j+1}$. The Hamiltonians $H_1$ and $H_2$ represent the PXP model acting on the left and the right halves of the system, respectively. The terms ${V}_1$ and ${V}_2$ determine the boundary conditions. We use ${V}_1 = {X}_1 {P}_2$ and ${V}_2 = {P}_{N-1} {X}_N$ for the open boundary conditions, and we use ${V}_1 = {P}_N {X}_1 {P}_2$ and ${V}_2 = {P}_{N-1} {X}_N {P}_1$ for the periodic boundary conditions. It is easy to check that the Hamiltonians $H_1$ and $H_2$ are not commutable to each other. 

The PXP model is originally derived as an effective model from the transverse Ising model. See the footnote 
\footnote{In the open boundary condition, the original transverse Ising model is given by \[ H= \sum_{i=1}^{N-1} J Z_i Z_{i+1} + \sum_{j=1}^N (2J - J (\delta_{j,1} + \delta_{j,N})) Z_i + \gamma_i X_i  \, , \] where $J > 0$ and $J \gg \gamma_i$ are imposed. We consider a constrained Hilbert space without any adjacent up states. When set $\gamma_i =0$, all eigenstates in this restricted Hilbert space are degenerate in their energies. Then, the transverse field term $\sum_i \gamma_i X_i$ is treated by the perturbation to get the PXP model. From this picture, the decomposition of the Hamiltonians (\ref{Ham L}) and (\ref{Ham R}) can be derived by applying the time-dependent transverse field. Namely, we define $\gamma_L=\gamma_i (i=1,\cdots , N/2 )$ and $\gamma_R=\gamma_i (i=N/2+1,\cdots , N)$, then we apply time-dependent fields as $(\gamma_L,\gamma_R)=(1,0)$ for $0\le t < T/2$, and $(\gamma_L,\gamma_R)=(0,1)$ for $T/2 \le t < T$.} 
for the physical implementation on the above-mentioned decomposition of Hamiltonians using the original transverse Ising model. 
In Rydberg-dressed atom experiments, the ground state $\state{g}_j$ and the Rydberg state $\state{r}_j$ trapped into a tweezer array at $j$ correspond to $\state{\downarrow}_j$ and $\state{\uparrow}_j$, respectively \cite{SchausRydbergIsing15,	LabuhnRydbergIsing16, bernien}. 
The PXP interaction is switched on and off by turning the laser on and off, which induces a Rabi oscillation between $\state{g}_j$ and $\state{r}_j$ \cite{bernien, YaoFSPT17,GrahamSuffman2019}.

As an indicator of a nonthermal (or thermal) state, we use the entanglement entropy $S_{N/2}:=-{\rm Tr}_1 \rho_1 \log \rho_1$, where $\rho_1$ is the reduced density matrix obtained by taking the partial trace with respect to the sites $j\in[N/2+1,N]$. Thermal states have a large amount of the entanglement entropy, which are linearly dependent on the system size $N$ \cite{Page:1993df, doi:10.1080/00018732.2016.1198134, Nakagawa2018, GroverPageCurve19}. If the entanglement entropy is small and independent of $N$, the state is an nonthermal state. Now, let us discuss the entanglement entropy as a function of the period $T$. We numerically calculate the entanglement entropies for all Floquet eigenstates. We use the open boundary condition with the system sizes $N=16$ and $20$. Fig.~\ref{tdepe} shows the $T$-dependence of the minimum value of the entanglement entropies among all the eigenstates; interestingly, a resonance-like phenomenon is observed. The inset shows a magnified plot around $T\sim 44.4$, where almost vanishing entanglement entropy is observed, indicating a scar state. An important question is whether this is the state intrinsic to the Floquet operator or simultaneous eigenstate of the static Hamiltonian. To address this question, we consider the entanglement entropies for all Floquet eigenstates at a fixed period $44.4$, and compare them with the entanglement entropies for the eigenstates of $H_1+H_2$. If the Floquet scars are the simultaneous eigenstates for $H_j \, (j=1,2)$, one will see the coincidence of entanglement entropies between them. In Fig.~\ref{fig entanglement}, we present them as a function of the expectation value of $H_1 + H_2$ for each eigenvalue. In the figure, the orange dots and the blue points represent the results for Floquet eigenstates and eigenstates of $H_1+H_2$, respectively. Almost all Floquet eigenstates are thermal, as indicated by large values of entanglement entropies. The values are close to the theoretical value estimated from the reduced density matrix with an infinite temperature, i.e., $S_{N/2}\sim 4.4$ 
\footnote{$ S_{N/2} = \ln |{\cal H}_{N/2}|-1/2$ where $|{\cal H}_{N/2}|$ is the number of Hilbert space for the size $N/2$. The subtracted term 1/2 always appears in a finite-size system \cite{Page:1993df}.}
. Remarkably, exceptions for four states are seen, which have zero entanglement entropies (two states are degenerated at energy $=0$, and only three points are thus visible in the figure); however, all eigenstates of $H_1+H_2$ have finite entanglement entropies. Thus, the Floquet scars observed here are not simultaneous eigenstates of the static Hamiltonian, which are the desired FMS. 

\sectionprl{Exact description of the FMS}
We make the above numerical indication rigorous in the following, by providing an explicit description of the many-body scar state. We eventually show the following form for the FMS $\state{\rm FS_{\alpha,\beta}}$ for the period $T= t_m :=2\sqrt{2} \pi m$ with a positive integer $m$: 
\begin{align}
	\state{\rm FS_{\alpha,\beta}}
	&=
	\state{\Phi_{1 , \,  \alpha} } \otimes
	\state{\downarrow}_{N/ 2}\state{\downarrow}_{{N/ 2}+1} \otimes
	\state{\Phi_{2,\, \beta} } \, , \label{FS state}
\end{align}
where we obtain four FMSs for $\alpha, \beta = \pm$, and the detailed expressions of $\state{\Phi_{1, \, \alpha}}$ and $\state{\Phi_{2, \, \alpha}}$ are given below. Even at this level, we can list several physically crucial aspects. First, the period in the inset of Fig.~\ref{fig entanglement} is consistent with the value $T=10\sqrt{2} \pi$ (i.e., $m=5$). Second, the entanglement entropy is exactly zero from the structure of the above-mentioned expression. Third, the above expressions are different from the Lin-Motrunich (LM) eigenstates for the static Hamiltonian $H_1+H_2$ \cite{LM}. The LM eigenstate is given with the matrix product form, which is clearly different from the above form. As explained below, the definition of the FMSs is that they are not the eigenstates of $H_j$, but the simultaneous eigenstates for the unitary operations $e^{-iH_1 t_m/2}$ and $e^{-iH_2 t_m/2}$.

We now consider the detailed expression for (\ref{FS state}). Although the LM eigenstates are not identical to the FMS, those are still beneficial in deriving the FMSs. We first focus on the left part that consists of the sites from $j=1$ to $j=N/2+1$. We note that the $Z_{N/2+1}$ and $H_1$ are commutable to each other. When we fix the state at $j=N/2+1$ to the down state, $H_1$ can be regarded as the PXP model of the system size $N/2+1$ with the open boundary condition. In this case, the LM eigenstates are given by the following matrix product form:
\begin{align}
	\state{\Gamma_{\alpha, \beta}} 
  & = \sum_{\{\sigma\}} v^T_\alpha B^{\sigma_1}C^{\sigma_2}\cdots B^{\sigma_{{N/ 2}-1}}C^{\sigma_{{N/ 2}}} v_\beta \nonumber \\
  & \times \state{\sigma_1\cdots \sigma_{N / 2 }} \otimes \state{ \downarrow}_{{N/ 2 }+1} \, , 
	  \label{Lin Motrunich solution}
\end{align}
where $v_{\pm} \equiv (1,\pm 1)^{T}$ and 
\begin{align}
	B^\downarrow 
	=
	\begin{pmatrix}
		1 & 0 & 0 \\
		0 & 1 & 0
	\end{pmatrix}\, ,\ &
	B^\uparrow 
	=
	\sqrt{2}
	\begin{pmatrix}
		0 & 0 & 0 \\
		1 & 0 & 1
	\end{pmatrix}\, , \\
	C^\downarrow
	=
	\begin{pmatrix}
		0 & -1 \\
		1 & 0 \\
		0 & 0
	\end{pmatrix} \, ,\ &
	C^\uparrow
	=
	\sqrt{2}
	\begin{pmatrix}
		1 & 0  \\
		0 & 0 \\
		-1 & 0
	\end{pmatrix} \, .
\end{align}
The eigenenergies are E=0 for $\state{\Gamma_{\pm, \pm}}$ and $E=\pm \sqrt{2}$ for $\state{\Gamma_{\pm, \mp}}$.

In addition, we make a new wave function by a linear transformation:
\begin{align}
	\state{\Psi_{1,\, \alpha}} &= {1\over \sqrt{2}}
				\left( \state{\Gamma_{\alpha ,+}}-\state{\Gamma_{\alpha,-}} \right) \, ,  ~~~\alpha = +,- \, . \label{leftrightm}
\end{align}
Through straightforward calculation, this state turns out to be identical to the following expression
\begin{align}
  \state{\Psi_{1, \alpha}} & = \state{\Phi_{1, \alpha} } \otimes \state{\downarrow
  }_{N/2}\state{\downarrow}_{{N/ 2}+1} \label{disentangle left} \, ,
\end{align}
where 
\begin{align}
	\state{\Phi_{1, \alpha}}	
&	= \sum_{\{\sigma \}} v^T_{\alpha} B^{\sigma_1}C^{\sigma_2}\cdots B^{\sigma_{{N/ 2}-1}} w_\rmL  \nonumber \\
& ~~~~~~~~~~~~ \times   \state{\sigma_1\cdots \sigma_{{N / 2}-1 }} \, , 
                             \label{phi L}
\end{align}
with the new vector $w_ \rmL = (\sqrt{2},0,0)^T$. It should be noted that the state \eqref{leftrightm} is a superposition of the two eigenstates with the energy $E=0$ and $\sqrt{2}$, and is therefore not the eigenstate of the Hamiltonian $H_1$. However, when we consider the unitary time evolution $e^{-it H_1}$ starting from this state, the wave function returns to the initial state with the time $t= \sqrt{2} \pi$, i.e., the two states $\state{\Psi_{1,\, \alpha}} \, (\alpha=\pm)$ are the eigenstates of $e^{-i H_1 t_m/2}$.

A similar analysis is performed for the unitary time-evolution $e^{-i H_2 t_m /2}$ by considering the site $j=N/2$ to $j=N$. Starting with fixing the state at the site $N/2$ to the down state, we eventually arrive at the following wave function: 
\begin{align}
  \state{\Psi_{2, \beta}} & = \state{\downarrow}_{N/ 2} \state{\downarrow}_{{N / 2}+1} \otimes \state{\Phi_{2 , \beta}}  \, ,   \label{disentangle right}
\end{align}
where $\beta = +,- $, and the function $\state{\Phi_{2 , \beta}}$ is given by 
\begin{align}
	\state{\Phi_{2, \beta} }	
	& = \sum_{\{\sigma\}} w_\rmR^T 
   C^{\sigma_{{N/ 2}+2}}B^{\sigma_{{N / 2}+3}}\cdots C^{\sigma_{N}} v_{\beta } \nonumber \\
  & ~~~~~~~~~~~~ \times  \state{\sigma_{N/2+2}\sigma_{N/2+3}\cdots \sigma_{N}}, \label{phi R}
\end{align}
with the vector $w_\rmR = (\sqrt{2},0,0)^T$. The states (\ref{disentangle right}) are the eigenstates for $e^{-i H_2 t_m/2}$.

Crucially, both $\state{\Phi_{1, \alpha }}$ and $\state{\Phi_{2, \beta }}$ contain the product states with the down states at the site $N/2$ and $N/2+1$, and thus, we can safely merge these states to obtain the desired expression (\ref{FS state}). From these derivations, one can see that our FMSs are not the eigenstates of the static Hamiltonian ${H}_j \, (j=1,2)$, but they are the simultaneous eigenstates for unitary operators $e^{-i H_1 t_m/2}$ and $e^{-i H_2 t_m/2}$. The rigorous proof of the FMS by showing the explicit Floquet eigenstate, and the underlying physical mechanism, are the main results in this paper.

\sectionprl{Periodic boundary condition and the Floquet-scar-engineering}
\begin{figure}
		\includegraphics[width= 0.45 \textwidth]{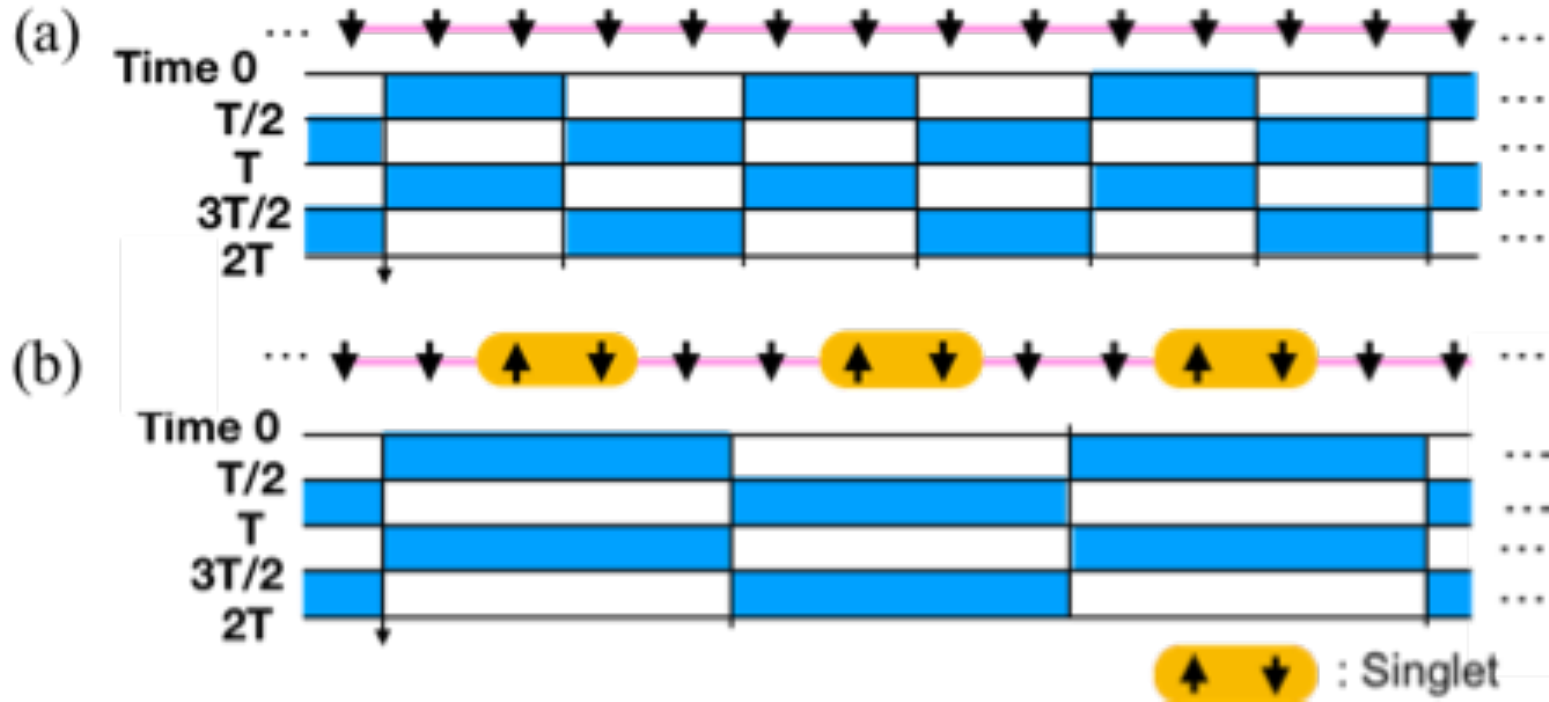}
		\caption{
		Schematic pictures of an example of the protocol \eqref{protocol evenodd 1} and \eqref{protocol evenodd 2}. 
		(a): Two consecutive interactions are turned on and off. Blue areas in the figure denote the periods when the interactions are turned on. Our solution \eqref{FS evenodd} reduces to the all-down state. 
		(b): Same as (a) for $n=4$, in which four consecutive interactions are turned on and off. Two down spins and the singlet state appear alternatively in the FMS. 		}
		\label{fig evenodd}
\end{figure}
In the case of the periodic boundary condition with the period $T=2\sqrt{2}\pi m$, one can readily obtain the FMS by following a similar procedure as above. In this case, we have only one scar state, which is given by 
\begin{align}
	\state{{\rm FS} } 
	=&
	\state{\downarrow}_{N}\state{\downarrow}_{1} 
	\otimes
	\state{\tilde{\Phi}_{2,{N/ 2}-1}} \non \\
	&\otimes \state{\downarrow_{N/ 2}}\state{\downarrow_{{N / 2}+1}} 
	\otimes
	\state{\tilde{\Phi}_{{N / 2}+2,N-1}} \, ,
	\label{FS periodic}	
\end{align}
where $\state{\tilde{\Phi}_{i,j}}$ is a pure state defined from the site $i$ to $j$: 
\begin{align}
	\state{\tilde{\Phi}_{i,j}}
	&=
	\sum_{\{\sigma\}} w_\rmR^T 
	C^{\sigma_{i}}B^{\sigma_{i+1}}\cdots B^{\sigma_{j}} w_L
	  \state{\sigma_i \cdots \sigma_{j}}.
\end{align}
The entanglement entropy for the subsystem consisting of the sites $i=1,\cdots, N/2$ is exactly zero from the structure. 

Having understood the underlying mechanism to have scar states, we now demonstrate that other systems which have scar states can be systematically engineered. We emphasize that many systems can be systematically constructed. We show an example of such applications below.
\begin{figure}
		\includegraphics[width= 0.45 \textwidth]{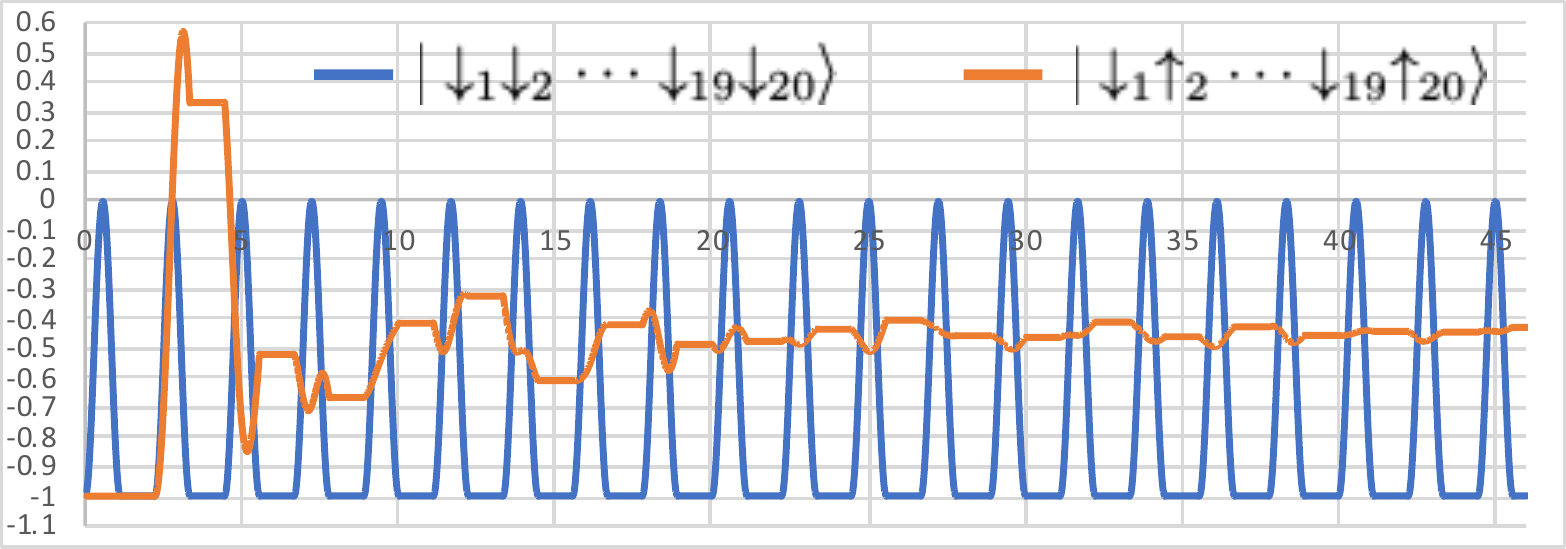}
		\caption{
		Time evolution of the FMS and ${\mathbb Z}_2$ state under the protocol \eqref{protocol evenodd 1} and \eqref{protocol evenodd 2} for $n=2$. The vertical axis is the magnetization at site 5 and the horizontal axis is time. The length of the chain is 20. 
		(Blue): The initial state is the all-down state, which is the FMS for $n=2$. Since this is the eigenstate of the Floquet operator, the spin perfectly returns to the down state $\state{\downarrow_5}$ in every period. 
		(Orange): The initial state is the ${\mathbb Z}_2$ state $\state{\downarrow_1 \uparrow_2 \cdots \downarrow_{19} \uparrow_{20} }$. This is not the FMS, and thus, the amplitude of the oscillation decays in time. 
		}
		\label{fig time evolution}
\end{figure}

We consider the system of size $N= 2 n\ell$, where $n$ and $\ell$ are integers. Then, we make a unitary time-evolution of each $n$ sites  by dividing the Hamiltonian as follows: 
\begin{align}
	{H}_1 
& =\sum_{k=0}^{\ell -1} \sum_{j=2nk+1  }^{n(2k+1)}
  {h}_j \, \label{protocol evenodd 1}\\
  ~~~~~{H}_2 
&=\sum_{k=0}^{\ell -1 } \sum_{j= n (2k+1) +1 }^{n (2k+2)}
	{h}_j \, , \label{protocol evenodd 2} 
\end{align}
where ${h}_j := {P}_{j-1} {X}_j {P}_{j+1}$ and we impose the periodic boundary condition or the open one. By following the same procedure as before, regardless of the boundary conditions one can find the exact scar state for the period $T=t_m$: 
\begin{align}
	\state{{\rm FS}}
	& =
	\bigotimes_{k=0}^{2 \ell -1}
	\state{\downarrow_{nk+1}}\otimes
	\state{\tilde{\Phi}_{nk+2,n(k+1) -1}}	
	\otimes \state{\downarrow_{n (k+1)}} \, . \label{FS evenodd}	
\end{align}
In this FMS, $\state{\tilde{\Phi}_{i,j}}$ and $\state{\downarrow\downarrow}$ appear alternatively. 
We note that the spins at the edges $j=1$ and $N$ are both down states for the open boundary condition. 
In \eqref{FS state} four FMS states do not have the down states at the edge. However, by superposing them we have the FMS which has $\state{\downarrow}_1$ and $\state{\downarrow}_N$. In the same way we make the edge state in \eqref{FS evenodd} the down states. 

In Fig.~\ref{fig evenodd}a and b, two simplest cases of the protocol \eqref{protocol evenodd 1}, \eqref{protocol evenodd 2}, and the FMS state \eqref{FS evenodd} are schematically illustrated. The upper figure in the inset is the case for $n=2$ and the lower one is for $n=4$. 
Interestingly, \eqref{FS evenodd} reduces to simple states; For $n=2$, it is an all-down state, because the term $\state{\tilde{\Phi}_{nk+2,n(k+1)-1}}$ vanishes. 
For $n=4$, $\state{\tilde{\Phi}_{nk+2,n(k+1)-1}}$ reduces to the singlet state. 
We note that for $n=2$ \eqref{FS evenodd} becomes the FMS for $T=\sqrt{2}\pi m$ with positive integer $m$, because \eqref{FS evenodd} is the superposition of $\state{\Gamma_{+,-}}$ and $\state{\Gamma_{-,+}}$, whose eigenenergies are  $\pm\sqrt{2}$. 
This is twice as frequent as the periods for $n\geq 4$, because \eqref{FS evenodd} for $n\geq 4$ is the superposition of $\state{\Gamma_{+,-}}$, $\state{\Gamma_{-,+}}$ $\state{\Gamma_{-,-}}$ and $\state{\Gamma_{+,+}}$, whose eigenenergies are  $\sqrt{2}$, $-\sqrt{2}$, 0 and 0, respectively. The detailed explanation is provided in the Supplementary materials. 
The advantage of the FMS is clearly observed in time evolution. 
In Fig.~\ref{fig time evolution}, numerical results of the time evolution starting from the FMS $\state{\downarrow \downarrow\cdots}$ (blue) and from the ${\mathbb Z}_2$ state $\state{\uparrow \downarrow \uparrow \downarrow\cdots}$ (orange) are shown. We impose the open boundary condition and take $T=\sqrt{2}\pi$ in our numerical calculation. 
Starting with the FMS we see perfect revival to the initial state, while the spin decays quickly for the ${\mathbb Z}_2$ state. 
The value of $\langle Z_5\rangle$ after the relaxation is around $-1/\sqrt{5}$, which is the ensemble average value at infinite temperature. We note that this is nonzero due to the Rydberg blockade. 

\sectionprl{Summary and perspective}
In this paper, we discussed the Floquet many-body scar states. We first classified the possible scar states into two classes. In the first class, the scar state is the simultaneous eigenstate for the Hamiltonian $H(t)$ for any $t$. This type is given by using the frustration-free Hamiltonian, such as the AKLT Hamiltonian. We focus on the second class, which has the scar state intrinsic to the Floquet operator (FMS). Our model consists of the PXP-type interactions without disorder. We exactly demonstrate that the FMSs certainly exist, by showing the explicit expressions of the eigenstates. The crucial mechanism of the FMSs discovered here is that the states are simultaneous eigenstates of different unitary operators, while they are not simultaneous eigenstates for the Hamiltonians. 
Another important feature is the absence of the conserved quantities in our system. Hence the mechanism presented in this paper should be different from those in the previous works where the system has conserved quantities \cite{Asmi2019}. 
All the Hamiltonians in this study can be implemented in a chain of Rydberg dressed alkali-metal atoms in principle \cite{bernien,YaoFSPT17,GrahamSuffman2019}. In particular, the protocols depicted in Fig.~\ref{fig evenodd} are the most feasible for experimental realization, because the FMS states reduce to simple states which can be readily prepared in experiments. It is an important future subject to observe the FMS in cold atoms experiments. 

\sectionprl{acknowledgments}
The authors thank W.~W.~Ho, H.~Levine, and H.~Katsura for useful discussions
and valuable comments. 
S.S was supported by JSPS Overseas Research Fellowships (201860254). 
T.K. was supported by the RIKEN Center for AIP and JSPS KAKENHI Grant No. 18K13475.
K.S. was supported by JSPS Grants-in-Aid for Scientific Research (JP16H02211).

\bibliography{Floquet_scar_arXiv}  

\appendix
\section{Two-sites protocol}
We show the results of the system where two consecutive sites are periodically driven. 
The two sites protocol is schematically shown in Fig.~3 in the main text	. 
We show that the FMSs are stabler than other states under the time evolution of this protocol. 
Also, we find that the two sites protocol is a somewhat special case which has more FMSs than other protocols in our paper have. The results are summarized as follows. 
As explained in the main part, the all-down state $\state{\downarrow\downarrow \cdots\downarrow}$ is a FMS of certain periods, $T={2n\pi /\sqrt{2}}$ where $n$ is an arbitrary integer number. 
However, this is not the only FMS in the two sites protocol. We find that there are exponentially-large number of other FMSs. 
We also show that FMSs exist in other periods. 

\subsection{Four-sites system}
In order to explicitly see the time evolution of each period, we consider the minimum model of four sites: 
\begin{align}
	{H}_{\rm 2 sites} \equiv P_1X_2P_3+P_2X_3P_4, \label{Hamiltonian 2sites}
\end{align}
where the Rabi oscillation is induced on site 2 and 3. 
The eigenenergies and eigenstates are given by 
\begin{align}
	&\state{0_1}=\state{\uparrow \downarrow \downarrow \uparrow} \label{ev1}\\
	&\state{0_2}=\state{\downarrow \uparrow \downarrow \downarrow}
	-\state{\downarrow \downarrow \uparrow \downarrow}\label{ev2}\\
	&\state{\sqrt{2}}=\sqrt{2}\state{\downarrow \downarrow \downarrow \downarrow}
	+\state{\downarrow \uparrow \downarrow \downarrow}
	+\state{\downarrow \downarrow \uparrow \downarrow}\label{ev3}\\
	&\state{-\sqrt{2}}=-\sqrt{2}\state{\downarrow \downarrow \downarrow \downarrow}
	+\state{\downarrow \uparrow \downarrow \downarrow}
	+\state{\downarrow \downarrow \uparrow \downarrow}\label{ev4}\\
	&\state{1_1}=\state{\uparrow \downarrow \downarrow \downarrow}
	+\state{\uparrow \downarrow \uparrow \downarrow}\label{ev5}\\
	&\state{1_2}=\state{\downarrow \downarrow \downarrow \uparrow}
	+\state{\downarrow \uparrow \downarrow \uparrow}\label{ev6}\\
	&\state{-1_1}=\state{\uparrow \downarrow \downarrow \downarrow}
	-\state{\uparrow \downarrow \uparrow \downarrow}\label{ev7}\\
	&\state{-1_2}=\state{\downarrow \downarrow \downarrow \uparrow}
	-\state{\downarrow \uparrow \downarrow \uparrow},\label{ev8}
\end{align}
where numbers of the states are the energy eigenvalues and the subscripts are the indices of degenerated states, e.g., $\state{0_1}$ and $\state{0_2}$ are two eigenstates of zero energy eigenvalue. 

We consider the unitary time evolution given by 
\begin{align}
U_{\rm 2 sites} \equiv e^{-i {H}_{\rm 2 sites} T},\label{unitary 2sites} 
\end{align}
and $T={2n\pi /\sqrt{2}}$ where $n$ is an integer. The states \eqref{ev1}-\eqref{ev4} do not acquire any phase after the time evolution \eqref{unitary 2sites}, while the others \eqref{ev5}-\eqref{ev8} do. 
Therefore, consider a superposition of the states \eqref{ev1}-\eqref{ev4}
\begin{align}
	\state{\Phi} = a\state{0_1}+b\state{0_2}+c\state{\sqrt{2}}+d\state{-\sqrt{2}},
\end{align}
where $a,b,c,d$ are any complex number. 
$\state{\Phi}$ does not experience dephasing under $U_{\rm 2 sites}$:
\begin{align}
	U_{\rm 2 sites}\state{\Phi}=\state{\Phi}. 
\end{align} 
$U_{\rm 2 sites}$ works as an identity operator for $\state{\Phi}$. 
Similarly, for $T={(2n+1)\pi /\sqrt{2}}$ where $n$ in an integer \eqref{ev3} and \eqref{ev4} acquire the same phase. Hence, any superposition of them are conserved by \eqref{unitary 2sites}. 

\subsection{FMS at $T={(2n+1)\pi /\sqrt{2}}$}
Given this mechanism of the minimum model, we can readily construct FMSs in the two sites protocol shown in Fig.~3 in the main text. Two Hamiltonians are applied to the system alternatively:
\begin{align}
	{H}_1 
& =\sum_{k=0}^{N/4-1} 
  {h}_{4k+1,4k+2} \, \label{protocol two sites 1}\\
  ~~~~~{H}_2 
&=\sum_{k=0}^{N/2 -1 }
  {h}_{4k+3,4k+4} \, \label{protocol two sites 2},
\end{align}
where ${h}_{j,j+1} := {P}_{j-1} {X}_j {P}_{j+1}+{P}_{j} {X}_{j+1} {P}_{j+2}$ and we impose the periodic boundary condition. 
When a quantum state is the product of the simultaneous eigenstate of \eqref{protocol two sites 1} and \eqref{protocol two sites 2}, it is a FMS. 
One solution exists at $T={(2n+1)\pi /\sqrt{2}}$ where $n$ is an integer number. 
With this period, any superposition of \eqref{ev3} and \eqref{ev4} returns to the same state after the time evolution in the Four sites system; $\state{\downarrow\downarrow\downarrow\downarrow}$ is such a state because $\state{\downarrow\downarrow\downarrow\downarrow}=(\state{\sqrt{2}}-\state{-\sqrt{2}})/2\sqrt{2}$. 
Since a longer chain described by \eqref{protocol two sites 1} and \eqref{protocol two sites 2} consists of the four-sites system, $\state{\downarrow_1\downarrow_2\cdots\downarrow_N}$ is a FMS for this protocol: 
\begin{align}
	U_F \state{\downarrow_1\downarrow_2\cdots\downarrow_N}=\state{\downarrow_1\downarrow_2\cdots\downarrow_N}.
	\label{unitary eigenstate}
\end{align}

\begin{figure}
		\includegraphics[width= 0.45 \textwidth]{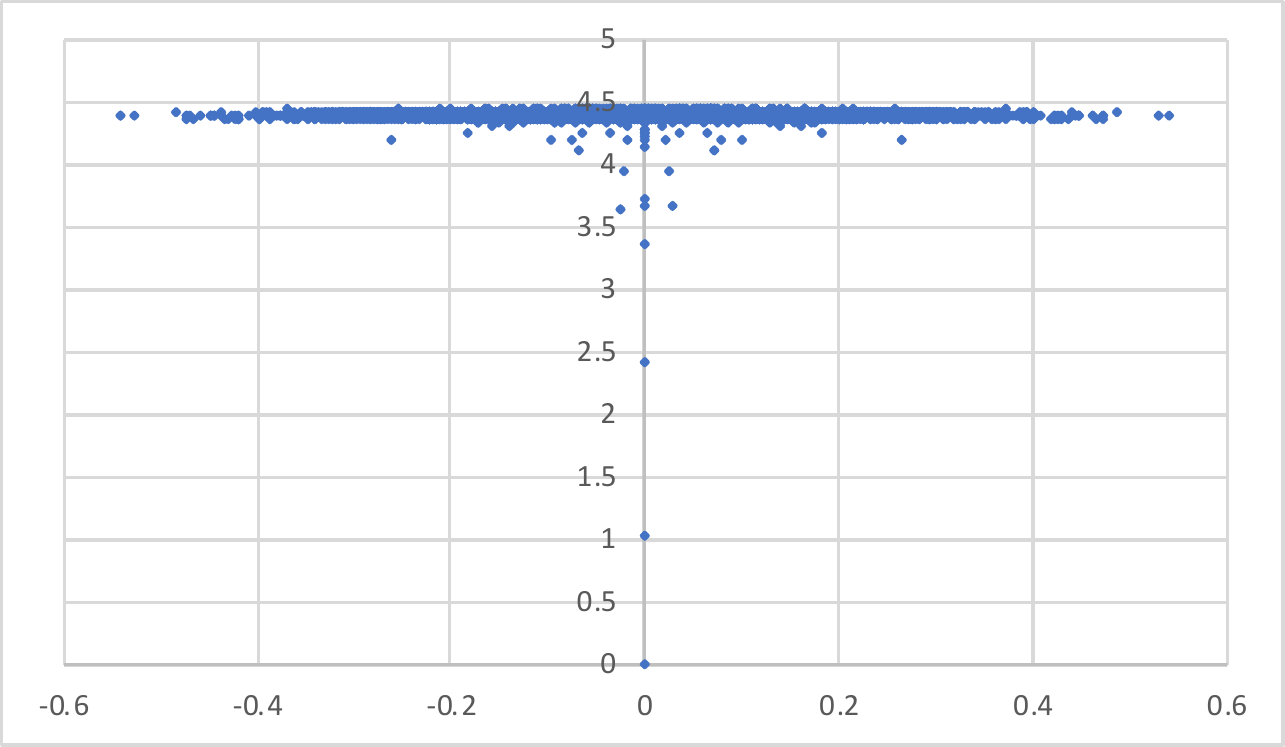}
		\caption{
		The entanglement entropy distribution of the Floquet eigenstates. The horizontal axis is $S_{N/2}$ and the vertical axis denotes the energy expectation with respect to the PXP Hamiltonian $H_1+H_2$.
	The Floquet period is $T_1=\pi/\sqrt{2}$, which is half of Fig.~\ref{fig EE 2sites}. 
	We see a FMS which has zero entanglement entropy at $\langle H_1+H_2 \rangle=0$ and several low-entanglement states. By contrast, other states satisfy the Floquet ETH. The EE of the Floquet eigenstates are concentrated around $S_{N/2}=4.4$, which is the average value at infinite temperature.} 
		\label{fig EE 2sites Half}
\end{figure}

In Fig.~\ref{fig EE 2sites Half}, we show the distribution of the entanglement entropy for $T=\pi/\sqrt{2}$. We see a scar state at $S_{N/2}=0$, while other states are distributed around 4.4, which is estimated from the density matrix at infinite temperature. 
At $\langle H_1+H_2 \rangle=0$, we also find several states which have small $S_{N/2}$. 
They are FMSs, e.g., for N=20 we can readily show that 
\begin{align}
	{1\over \sqrt{2}} (\state{OROLOROLOR}+\state{OLOROLOROL})
\end{align}
is a FMS whose Floquet eigenvalue is $\pi$, 
where we introduce notations
\begin{align}
	&\state{O}_j\equiv\state{\downarrow_{2j-1} \downarrow_{2j}}\\
	&\state{R}_j\equiv\state{\downarrow_{2j-1} \uparrow_{2j}}\\
	&\state{L}_j\equiv\state{\uparrow_{2j-1} \downarrow_{2j}}. 
\end{align}

The time evolution of the all-down state in this protocol is shown in Fig.~3 in the main text. We see the persistent oscillation in the whole periods as is expected for the FMS. 
By contrast, the spin decays rapidly in ${\mathbb Z}_2$ state in Fig.~3 in the main text. 

\subsection{Another FMS at $T=n\pi$}
Another solution of the FMS is found at $T={n\pi}$ where $n$ is an integer. 
The corresponding FMSs are constructed by superposing \eqref{ev5}-\eqref{ev8}. 
More precisely, $\state{\downarrow\uparrow\downarrow\uparrow}$ and $\state{\uparrow\downarrow\uparrow\downarrow}$ return to the same state in these periods. 
Therefore, $\state{\downarrow_1\uparrow_2\cdots\uparrow_{N-1}\uparrow_N}$ and 
$\state{\uparrow_1\downarrow_2\cdots\uparrow_{N-1}\downarrow_N}$ are FMSs for the protocol \eqref{protocol two sites 1} and \eqref{protocol two sites 2} in these periods.

\subsection{Exponentially many FMSs at $T={2n\pi /\sqrt{2}}$}
When $T={2n\pi/\sqrt{2}}$ for $n$ being an integer, the all-down state is again the eigenstate of the Floquet operator, \eqref{unitary eigenstate}. 
We show that exponentially large number of the Floquet eigenstate violate the Floquet ETH. 

Returning to the four-sites system, by superposing the states \eqref{ev1}-\eqref{ev4} we can make four product states, 
$	\state{\uparrow \downarrow \downarrow \uparrow},
	\state{\downarrow \uparrow \downarrow \downarrow},
	\state{\downarrow \downarrow \uparrow \downarrow},
	\state{\downarrow \downarrow \downarrow \downarrow}$.
Hence, if a product state is either of these four states for all ${h}_{4k+1,4k+2}$, this state is a FMS. 
More precisely, we can show that the configuration $\state{L}_i \state{O}_{i+1}\cdots \state{O}_{j-1}\state{R}_{j}$ is always conserved by $\hat{U}_F$, because we cannot flip the left edge state $\state{L}_i$ and the right edge state $\state{R}_{j}$. 
Hence, product states in which $\state{L}_i \state{O}_{i+1}\cdots \state{O}_{j-1}\state{R}_{j}$ and $\state{O}$ appear alternatively become the eigenstate of $\hat{U}_F$. 
For example, 
\begin{align}
\state{OLORO}
=
\state{\downarrow_1\downarrow_2\uparrow_3\downarrow_4\downarrow_5\downarrow_6\downarrow_7\uparrow_8\downarrow_9\downarrow_{10}} \label{example FMS1}\\
\state{LROLR}
=\state{\uparrow_1\downarrow_2\downarrow_3\uparrow_4\downarrow_5\downarrow_6\uparrow_7\downarrow_8\downarrow_9\uparrow_{10}}\label{example FMS2}
\end{align}
satisfy this condition. 
Since the number of such FMSs can be counted in a similar manner to how we count the dimension of PXP model, the number of the FMSs is an exponential of $N$. 

\begin{figure}
		\includegraphics[width= 0.45 \textwidth]{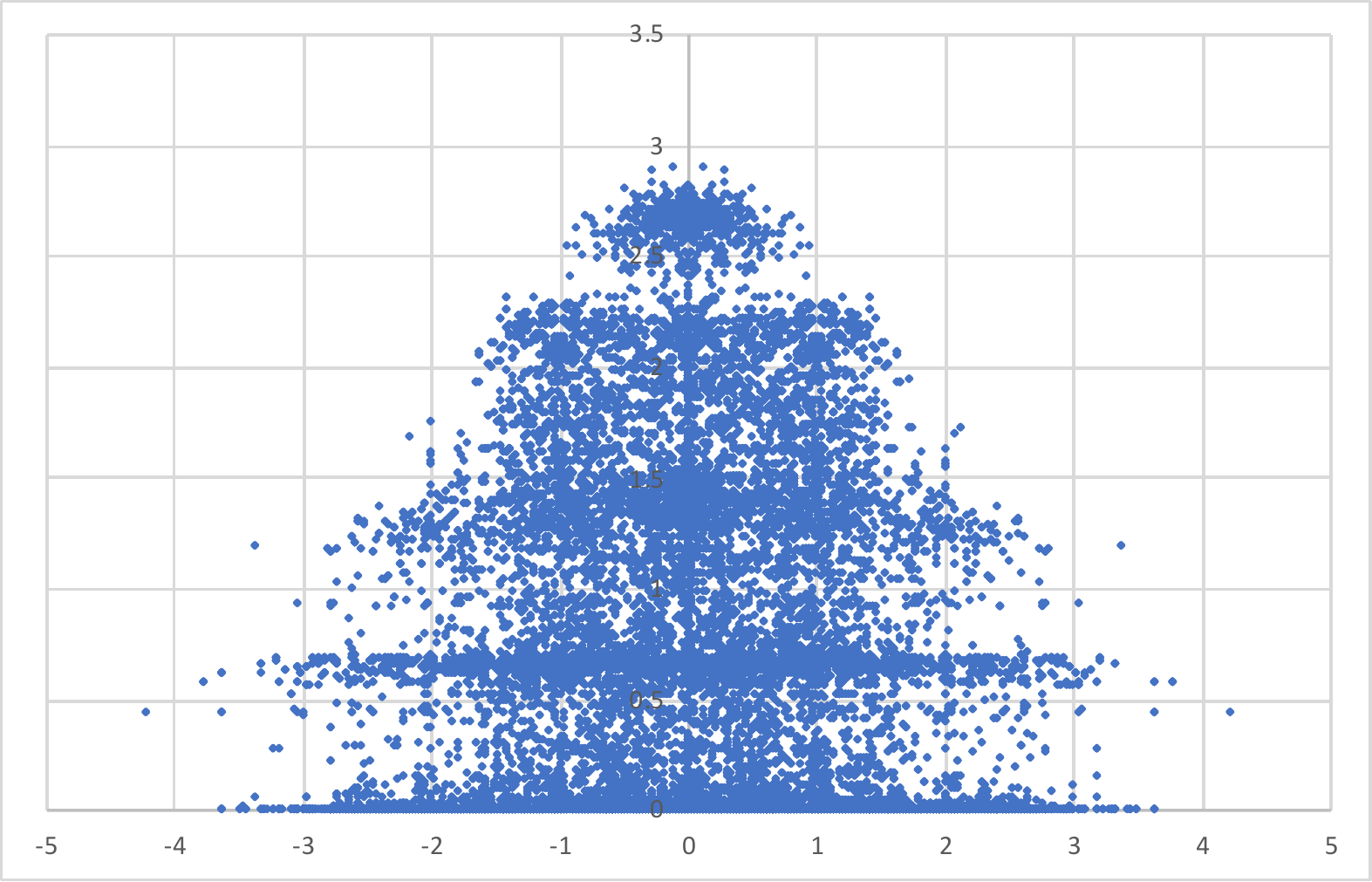}
		\caption{
	Same as Fig.~\ref{fig EE 2sites Half} with the Floquet period is $T_1=2\pi/\sqrt{2}$.
 We see exponentially large number of the Floquet eigenstates which have zero entanglement entropy. All of them are the FMSs.}
		\label{fig EE 2sites}
\end{figure}

In Fig.~\ref{fig EE 2sites}, the entanglement entropy of all the Floquet eigenstates in this model is shown. Exponentially large number of the Floquet eigenstate are located on the horizontal axis. They are the FMSs such as \eqref{example FMS1} and \eqref{example FMS2}. 
However, we also see many other Floquet eigenstates have lower entanglement entropy, e.g., large number of them are concentrated around $S_{N/2}\simeq1$. 


\end{document}